\documentclass[graybox]{svmult}

\usepackage{helvet}         
\usepackage{courier}        
\usepackage{type1cm}        
\usepackage{makeidx}         
\usepackage{graphicx}        
\usepackage{multicol}        
\usepackage[bottom]{footmisc}

\usepackage{amssymb,latexsym,amsmath,amsbsy}
\usepackage{cite}
\usepackage{stmaryrd}  
\usepackage{arydshln}  
\usepackage{mathdots}  
\def\nn{\nonumber}
\def\deg{\mathop{\rm deg}\nolimits}

\def\qdots{\mathinner{\mkern1mu\raise1pt\vbox{\kern7pt\hbox{.}}\mkern2mu \raise4pt\hbox{.}\mkern2mu\raise7pt\hbox{.}\mkern1mu}}
\def\Z{{\mathbb Z}}

\def\gl{\mathfrak{gl}}
\def\ssl{\mathfrak{sl}}

\def\g{\mathfrak{g}}

\def\so{\mathfrak{so}}
\def\sp{\mathfrak{sp}}

\def\lb{\llbracket}
\def\rb{\rrbracket}

\def\beq{\begin{equation}}
\def\eeq{\end{equation}}
\DeclareMathOperator{\Tr}{Tr}
\DeclareMathOperator{\End}{End}
\DeclareMathOperator{\spn}{span}

\begin{document}

\title*{Matrix structure of classical ${\mathbb Z}_2 \times {\mathbb Z}_2$ graded Lie algebras}
\titlerunning{Classical ${\mathbb Z}_2 \times {\mathbb Z}_2$ graded Lie algebras}
\author{N.I.\ Stoilova and J.\ Van der Jeugt}
\institute{N.I.~Stoilova \at Institute for Nuclear Research and Nuclear Energy, 
Bulgarian Academy of Sciences,
Boul.\ Tsarigradsko Chaussee 72, 1784 Sofia, Bulgaria, \email{stoilova@inrne.bas.bg}
\and J.\ Van der Jeugt \at Department of Applied Mathematics, Computer Science and Statistics, Ghent University, 
Krijgslaan 281-S9, B-9000 Gent, Belgium, \email{Joris.VanderJeugt@UGent.be}}

\maketitle

\abstract{
A ${\mathbb Z}_2\times{\mathbb Z}_2$-graded Lie algebra $\mathfrak g$ is a ${\mathbb Z}_2\times{\mathbb Z}_2$-graded algebra $\mathfrak g$ with a bracket $\llbracket \cdot , \cdot \rrbracket$ that satisfies certain graded versions of the symmetry and Jacobi identity. In particular, despite the common terminology, $\mathfrak g$ is not a Lie algebra.
We construct classes of ${\mathbb Z}_2\times{\mathbb Z}_2$-graded Lie algebras corresponding to the classical Lie algebras, in terms of their defining matrices.
For the ${\mathbb Z}_2\times{\mathbb Z}_2$-graded Lie algebra of type $A$, the construction coincides with the previously known class.
For the ${\mathbb Z}_2\times{\mathbb Z}_2$-graded Lie algebra of type $B$, $C$ and $D$ our construction is new and gives rise to interesting defining matrices closely related to the classical ones but undoubtedly different.
We also give some examples and possible applications to parastatistics.
}

\section{Introduction}
\label{sec:1}

In physics, relations between physical operators $x$ and $y$ are commonly expressed in terms of commutators $[x,y]$ and anticommutators $\{x,y\}$.
Starting from an associative algebra, the bracket {$[x,y]=xy-yx$} leads to the structure of a {Lie algebra}. 
And starting from a $\Z_2$-graded associative algebra, the bracket {$\lb x, y \rb=xy - (-1)^{\xi \eta} yx$} (where $\xi$ is the degree of $x$ and $\eta$ is the degree of $y$) leads to the structure of a {Lie superalgebra}.
So why should one go beyond these structures, and study $\Z_2\times\Z_2$-graded Lie algebras or $\Z_2\times\Z_2$-graded Lie superalgebras?
The answer lies in the reformulation of the product order of three operators.
For {three elements $x,y,z$} in an associative algebra, the twelve terms of the trivial product identity
\begin{eqnarray}
& (xyz+yzx+zxy+zyx+yxz+xzy)\nn\\
& \qquad-(xyz+yzx+zxy+zyx+yxz+xzy)=0
\end{eqnarray}
can be rewritten by means of commutators and anticommutators in (essentially) four ways~\cite{YJ}:
\begin{align*}
\hbox{(a)}\qquad & [x,[y,z]]+[y,[z,x]]+[z,[x,y]]=0, \\
\hbox{(b)}\qquad & [x,\{y,z\}]+[y,\{z,x\}]+[z,\{x,y\}]=0, \\
\hbox{(c)}\qquad & [x,\{y,z\}]+\{y,[z,x]\}-\{z,[x,y]\}=0, \\
\hbox{(d)}\qquad & [x,[y,z]]+\{y,\{z,x\}\}-\{z,\{x,y\}\}=0. 
\end{align*}
Clearly, (a) corresponds to the Jacobi identity for a Lie algebra (LA); (a)--(c) appears as the Jacobi identity for a Lie superalgebra (LSA); but (d) can appear only as the Jacobi identity for a $\Z_2\times\Z_2$-graded Lie algebra or a $\Z_2\times\Z_2$-graded Lie superalgebra.

The definition of a $\Z_2\times\Z_2$-graded Lie algebra ({$\Z_2^2$-GLA}) or a $\Z_2\times\Z_2$-graded Lie superalgebras ({$\Z_2^2$-GLSA}) goes back to Rittenberg and Wyler~\cite{Rit1,Rit2}.
Let $\g$ be a $\Z_2\times\Z_2$-graded linear space:
\begin{equation}
\g=\bigoplus_{\boldsymbol{a}} \g_{\boldsymbol{a}} =
\g_{(0,0)} \oplus \g_{(0,1)} \oplus \g_{(1,0)} \oplus \g_{(1,1)}, 
\label{ggrading}
\end{equation}
where $\boldsymbol{a}=(a_1,a_2) \in \Z_2\times\Z_2$.
The relations that should hold are often written in terms of homogeneous elements, such as
$x_{\boldsymbol{a}}\in \g_{\boldsymbol{a}}$, which is of degree $\deg x_{\boldsymbol{a}}= \boldsymbol{a}$, and extended by linearity to all elements of $\g$.

The structure $(\g,\lb\cdot,\cdot\rb)$ with $\lb\cdot,\cdot\rb$ a bilinear operation is a $\Z_2\times\Z_2$-graded Lie algebra if
\begin{align}
& \lb x_{\boldsymbol{a}}, y_{\boldsymbol{b}} \rb \in \g_{\boldsymbol{a}+\boldsymbol{b}}, \label{ZZgrading} \\
& \lb x_{\boldsymbol{a}}, y_{\boldsymbol{b}} \rb = -(-1)^{\boldsymbol{a}\cdot\boldsymbol{b}} 
\lb y_{\boldsymbol{b}}, x_{\boldsymbol{a}} \rb, \\
& \lb x_{\boldsymbol{a}}, \lb y_{\boldsymbol{b}}, z_{\boldsymbol{c}}\rb \rb =
\lb \lb x_{\boldsymbol{a}}, y_{\boldsymbol{b}}\rb , z_{\boldsymbol{c}} \rb +
(-1)^{\boldsymbol{a}\cdot\boldsymbol{b}} \lb y_{\boldsymbol{b}}, \lb x_{\boldsymbol{a}}, z_{\boldsymbol{c}}\rb \rb,
\end{align} 
where
\begin{equation}
\boldsymbol{a}+\boldsymbol{b}=(a_1+b_1,a_2+b_2), \qquad
{\boldsymbol{a}\cdot\boldsymbol{b} = a_1b_2-a_2b_1}.
\label{ab}
\end{equation}
Note the peculiar product in~\eqref{ab}. 
If the product in~\eqref{ab} is replaced by $\boldsymbol{a}\cdot\boldsymbol{b} = a_1b_1+a_2b_2$, then the above is the definition of a $\Z_2\times\Z_2$-graded Lie superalgebra~\cite{Rit2}.

So in general, these algebras are not Lie algebras nor Lie superalgebras. 
The terminology is therefore slightly misleading.
But since these terms have become common now in literature, we shall also use them.

Compared to the essential role of Lie algebras and Lie superalgebras, 
the $\Z_2\times\Z_2$-graded Lie (super)algebras received for many years little attention in theoretical 
and mathematical physics~\cite{LR1978,Vasiliev1985,JYW1987}.
Only in recent years there is renewed interest in $\Z_2\times\Z_2$-graded Lie (super)algebras.
For example, such structures have appeared in symmetries of L\'evy–Leblond equations~\cite{Aizawa1,Aizawa3},
in graded (quantum) mechanics and quantization~\cite{Bruce2,AMD2020,Aizawa4,Aizawa5,Quesne2021},
and in $\Z_2\times\Z_2$-graded two-dimensional models~\cite{Bruce1,Toppan1,Bruce3}.
In particular $\Z_2\times\Z_2$-graded Lie algebras and superalgebras have been recognized in 
parastatistics~\cite{Tolstoy2014,SV2018} and in the description of parabosons and parafermions~\cite{Toppan2,Toppan3}.

In~\cite{SV2023} several classes of $\Z_2\times\Z_2$-graded Lie algebras corresponding to classical Lie algebras were constructed, in terms of defining matrices.
In the present contribution, we will recall some of these classes, but also characterize these classes of matrix algebras in a new way, by using the notion of graded transpose.

\section{$\Z_2\times\Z_2$-graded Lie algebras}
\label{sec:2}

Following the definition given in the previous section, it is clear that $\g_{(0,0)}$ is a Lie subalgebra of the $\Z_2^2$-GLA $\g$ and that $\g_{(0,1)}$,  $\g_{(1,0)}$ and $\g_{(1,1)}$ are $\g_{(0,0)}$-modules.
It will be useful to list the brackets (commutators or anticommutators) among the subspaces more explicitly:
\begin{equation}
[\g_{(0,0)}, \g_{\boldsymbol{a}}]\subset \g_{\boldsymbol{a}}, \quad 
[\g_{\boldsymbol{a}}, \g_{\boldsymbol{a}}]\subset \g_{(0,0)}, \qquad \boldsymbol{a}\in\Z_2\times\Z_2
\end{equation}
and
\begin{equation}
\{\g_{\boldsymbol{a}}, \g_{\boldsymbol{b}}\}\subset \g_{\boldsymbol{c}}
\end{equation}
if $\boldsymbol{a}$, $\boldsymbol{b}$ and $\boldsymbol{c}$ are mutually distinct elements of $\{ (1,0),(0,1),(1,1)\}$.
Thus if $\g=\g_{(0,0)} \oplus \g_{(0,1)} \oplus \g_{(1,0)} \oplus \g_{(1,1)}$ is a $\Z_2^2$-GLA, then any permutation of the last three subspaces maps $\g$ into another $\Z_2^2$-GLA.
Such mappings are ``trivial permutation transformations'' of $\g$.
 
The three subspaces $\g_{(0,1)}$,  $\g_{(1,0)}$ and $\g_{(1,1)}$ are therefore on an equal footing. 
In order to exclude ordinary Lie algebras or Lie superalgebras, we can assume that at least two of these three subspaces are nontrivial.
And following the previous remark, we shall impose the condition that $\g$ is generated by $\g_{(1,0)}\oplus \g_{(0,1)}$ (both nontrivial). 
Then one can deduce from the Jacobi identity that
\begin{equation}
\g_{(0,0)} = [\g_{(1,0)},\g_{(1,0)}] + [\g_{(0,1)},\g_{(0,1)}] \qquad\hbox{and}\qquad
\g_{(1,1)} = \{\g_{(1,0)},\g_{(0,1)}\}.
\label{g10g01}
\end{equation}

As usual, one can construct a $\Z_2^2$-GLA from any $\Z_2\times\Z_2$ graded associative algebra.
Indeed, let $\g$ be an associative algebra with a product $x\cdot y$, and suppose $\g$ has a $\Z_2\times\Z_2$-grading of the form~\eqref{ggrading} that is compatible with the product, i.e.\ $x_{\boldsymbol{a}} \cdot y_{\boldsymbol{b}} \in \g_{\boldsymbol{a}+\boldsymbol{b}}$.
Then it is easy to verify that the following bracket turns $\g$ into a $\Z_2^2$-GLA:
\begin{equation}
\lb x_{\boldsymbol{a}} , y_{\boldsymbol{b}}\rb = x_{\boldsymbol{a}} \cdot y_{\boldsymbol{b}}- 
(-1)^{\boldsymbol{a}\cdot\boldsymbol{b}} y_{\boldsymbol{b}} \cdot x_{\boldsymbol{a}}\ .
\label{ZZbracket}
\end{equation}

Concretely, let $V$ be a $\Z_2\times \Z_2$-graded linear space of dimension $n$:
{$V=V_{(0,0)} \oplus V_{(0,1)} \oplus V_{(1,0)} \oplus V_{(1,1)}$},
with subspaces of dimension $p,q,r$ and $s$ respectively, where $p+q+r+s=n$.
{$\End(V)$ is then a $\Z_2\times \Z_2$-graded associative algebra, 
and by the previous property it is turned into a $\Z_2^2$-GLA by the bracket $\lb\cdot,\cdot\rb$ of~\eqref{ZZbracket}. 
Let us denoted this algebra by $\gl_{p,q,r,s}(n)$.}
In matrix form, the elements are written as:
\begin{equation}
\begin{array}{c c}
    {\begin{array} {@{} c c  cc @{}} \ \ p\ \ \ & \ \ q\ \ \ & \ \ r\ \ \ & \ \ s \ \ \end{array} } & {} \\  
    \left(\begin{array}{cccc} 
a_{(0,0)} & a_{(0,1)} & a_{(1,0)} & a_{(1,1)} \\[1mm] 
b_{(0,1)} & b_{(0,0)} & b_{(1,1)} & b_{(1,0)} \\[1mm] 
c_{(1,0)} & c_{(1,1)} & c_{(0,0)} & c_{(0,1)} \\[1mm] 
d_{(1,1)} & d_{(1,0)} & d_{(0,1)} & d_{(0,0)} 
    \end{array}\right)
 & \hspace{-2mm} 
		{\begin{array}{l}
     p \\[1mm]  q \\[1mm]	r \\[1mm] s
    \end{array} } \\ 
  \end{array} 
\label{ZZsl}
\end{equation}
The indices of the matrix blocks refer to the $\Z_2\times\Z_2$-grading, and the size of the blocks is indicated in the lines above and to the right of the matrix.

One can check that $\Tr \lb A,B \rb =0$, where $\Tr$ is the ordinary trace, hence {$\g=\ssl_{p,q,r,s}(n)$ is defined as the subalgebra of traceless elements of $\gl_{p,q,r,s}(n)$}.
The dimensions of the subspaces of $\g$ are given by
\[
\begin{array}{ll}
\g_{(0,0)}\quad & p^2+q^2+r^2+s^2-1\\
\g_{(0,1)}\quad & 2pq+2rs\\ 
\g_{(1,0)}\quad & 2pr+2qs\\
\g_{(1,1)}\quad & 2qr+2ps
\end{array}
\]

Note that the underlying vector space is just the space of traceless matrices.
For the associative matrix algebra of traceless matrices, all $\Z_2\times\Z_2$-gradings have been classified~\cite[Example~2.30]{Elduque}, 
and these correspond to the gradings~\eqref{ZZsl}.
Using the bracket~\eqref{ZZbracket} (checking closure in the same vector space, and yielding the complete vector space) 
one obtains also in this way the $\Z_2^2$-GLA's $\ssl_{p,q,r,s}(n)$.
This class of $\Z_2^2$-GLA's is not new, and has been known for a long time. 
It appears already in~\cite{Rit2} under the name $sl(p,q,r,s)$.

\section{$\Z_2^2$-GLA's as subalgebras of $\ssl_{p,q,r,s}(n)$}
\label{sec:3}

A crucial notion to be introduced here is that of graded transpose.
Let $A$ be a homogeneous element of $\ssl_{p,q,r,s}(n) \subset \End(V)$ of degree $\boldsymbol{a}\in\Z_2\times\Z_2$.
Let $V^*$ be the vector space dual to $V$, inheriting the $\Z_2\times\Z_2$-grading from $V$, 
and denote the natural pairing of $V$ and $V^*$ by $\langle\cdot,\cdot\rangle$.
Then $A^* \in \End(V^*)$ is determined by:
\begin{equation}
\langle A^* y_{\boldsymbol{b}},x\rangle= (-1)^{{\boldsymbol{a}}\cdot{\boldsymbol{b}}} \langle y_{\boldsymbol{b}}, Ax \rangle,
\qquad \forall y_{\boldsymbol{b}}\in V^*_{\boldsymbol{b}}, \forall x\in V.
\label{defT}
\end{equation}
Clearly, this is extended by linearity to all elements of $\ssl_{p,q,r,s}(n)$.
In matrix form, this yields the {$\Z_2\times\Z_2$-graded transpose $A^T$} of $A$:
\begin{equation}
A=\left(\begin{array}{cccc} 
a_{(0,0)} & a_{(0,1)} & a_{(1,0)} & a_{(1,1)} \\ 
b_{(0,1)} & b_{(0,0)} & b_{(1,1)} & b_{(1,0)} \\ 
c_{(1,0)} & c_{(1,1)} & c_{(0,0)} & c_{(0,1)} \\ 
d_{(1,1)} & d_{(1,0)} & d_{(0,1)} & d_{(0,0)} 
    \end{array}\right),
A^T=\left(\begin{array}{cccc} 
a_{(0,0)}^t & b_{(0,1)}^t & c_{(1,0)}^t & d_{(1,1)}^t \\ 
a_{(0,1)}^t & b_{(0,0)}^t & -c_{(1,1)}^t & -d_{(1,0)}^t \\ 
a_{(1,0)}^t & -b_{(1,1)}^t & c_{(0,0)}^t & -d_{(0,1)}^t \\ 
a_{(1,1)}^t & -b_{(1,0)}^t & -c_{(0,1)}^t & d_{(0,0)}^t 
    \end{array}\right),
\label{AT}		
\end{equation}
where $a^t$ denotes the ordinary matrix transpose.
It is not difficult to check (case by case, according to the $\Z_2\times\Z_2$-grading) that the graded transpose of matrices satisfies
\begin{equation}
(AB)^T = (-1)^{{\boldsymbol{a}}\cdot{\boldsymbol{b}}} B^T A^T,
\label{ABT}
\end{equation}
where the sign is determined by~\eqref{ab}.

This notion of graded transpose allows us to determine certain classes of $\Z_2^2$-GLA as subalgebras of $\ssl_{p,q,r,s}(n)$.
Denote
\begin{equation}
\so_{p,q,r,s}(n) = \{A\in \ssl_{p,q,r,s}(n) \ |\  A^T+A=0 \}.
\label{defso}
\end{equation}
If $A,B \in \so_{p,q,r,s}(n)$, then
\begin{align*}
\lb A,B\rb^T &= (AB-(-1)^{{\boldsymbol{a}}\cdot{\boldsymbol{b}}}BA)^T\\
&=(-1)^{{\boldsymbol{a}}\cdot{\boldsymbol{b}}} B^T A^T-A^TB^T = (-1)^{{\boldsymbol{a}}\cdot{\boldsymbol{b}}} B A-AB=-\lb A,B\rb,
\end{align*}
thus \eqref{defso} is closed under the bracket $\lb\cdot,\cdot\rb$ and forms a $\Z_2^2$-graded Lie subalgebra of $\ssl_{p,q,r,s}(n)$.
The matrices of $\so_{p,q,r,s}(n)$ are of the following form:
\begin{equation}
\begin{array}{c c}
    {\begin{array} {@{} c c  cc @{}} \ \ p\ \ \ & \ \ q\ \ \ & \ \ r\ \ \ & \ \ s \ \ \end{array}} & {} \\  
    \left(\begin{array}{cccc} 
a_{(0,0)} & a_{(0,1)} & a_{(1,0)} & a_{(1,1)} \\[1mm] 
-a_{(0,1)}^t & b_{(0,0)} & b_{(1,1)} & b_{(1,0)} \\[1mm] 
-a_{(1,0)}^t & b_{(1,1)}^t & c_{(0,0)} & c_{(0,1)} \\[1mm] 
-a_{(1,1)}^t & b_{(1,0)}^t & c_{(0,1)}^t & d_{(0,0)} 
    \end{array}\right) 
 & \hspace{-2mm}	{\begin{array}{l}
     p \\[1mm]  q \\[1mm]	r \\[1mm] s
    \end{array} }\\ 
  \end{array} 
\label{matrixso}
\end{equation}
where $a_{(0,0)}$, $b_{(0,0)}$, $c_{(0,0)}$ and $d_{(0,0)}$ are antisymmetric matrices.

In order to continue in the future with more structural properties of $\Z_2^2$-GLA's (roots, root space decomposition, etc.), 
it would be interesting to identify a Cartan subalgebra of \eqref{matrixso}.
The classical choice (as for the Lie algebra $\so(n)$ of antisymmetric matrices) would lead here to a nonabelian algebra (since the Cartan subalgebra elements would not all be elements of $\so_{p,q,r,s}(n)_{(0,0)}$).
Hence we will construct some other subalgebras of $\ssl_{p,q,r,s}(n)$, with a matrix form that does exhibit a Cartan subalgebra consisting of diagonal matrices.

For classical LA's of type $B$, $C$, $D$, the appropriate matrix form exhibiting a Cartan subalgebra of diagonal matrices is given by the following block matrices~\cite{Fulton,Humphreys}:
\begin{equation}
\begin{array}{lll}
{ \begin{array}{ll} G=\so(2n+1) \\ (\dim G=2n^2+n)\end{array} }
& \begin{array}{c c}
    {\begin{array} {@{} c c c @{}} \ \ n\  & \ \ n\ \ & 1\ \end{array}} & {} \\  
    \left(
       \begin{array}{@{} c  c  c @{}}
        a & b & c\\[0.5mm] 
        d & -a^t & e\\[0.5mm]
				-e^t & -c^t & 0
       \end{array}
    \right)  & \hspace{-2mm}
		{\begin{array}{c}
     n \\[0.5mm]  n \\[0.5mm]	1 
    \end{array} } 
   \end{array} \qquad
& \hbox{$b$ and $d$ antisymmetric} ; \\[8mm]
{\begin{array}{ll} G=\sp(2n) \\ (\dim G=2n^2+n) \end{array}} 
& \begin{array}{c c}
 {\begin{array} {@{} c c @{}} \, n\  & \, n\  \end{array} } & {} \\  
    \left(
       \begin{array}{@{} c   c @{}}
        a & b \\[0.5mm] 
        c & -a^t
       \end{array}
    \right)  & \hspace{-2mm}
		{\begin{array}{c}
     n \\[0.5mm]  n 
    \end{array} } 
  \end{array} 
& \hbox{$b$ and $c$ symmetric} ;\\[8mm]
{\begin{array}{ll} G=\so(2n) \\ (\dim G=2n^2-n) \end{array}}
 & \begin{array}{c c}
   { \begin{array} {@{} c c @{}} \, n\  & \, n\  \end{array} } & {} \\  
    \left(
       \begin{array}{@{} c   c @{}}
        a & b \\[0.5mm] 
        c & -a^t
       \end{array}
    \right)  & \hspace{-2mm}
		{\begin{array}{c}
     n \\[0.5mm]  n 
    \end{array} }
  \end{array} 
&\hbox{$b$ and $c$ antisymmetric} .
\end{array}
\label{LA}
\end{equation}

In~\cite{SV2023}, we followed a procedure to construct such classes of $\Z_2^2$-GLA's corresponding to classical LA's. 
That procedure uses in fact the classification of so-called 5-gradings of classical LA's as determined in~\cite{SV2005LA}. 
We will not repeat this procedure here, but rather put some of the results of~\cite{SV2023} in the framework of the earlier introduced graded transpose.

The first class of $\Z^2_2$-GLA's we list here are those of type $C$. 
The $\Z_2^2$-GLA $\g=\sp_{p}(2n)$ consists of all matrices of the following block form:
\begin{equation}
\begin{array}{c c}
   { \begin{array} {@{} c c c c @{}} \ \ \ \ p\ \ & \ \ n-p\ \ & \ p\ \ \ \ & \ \ n-p\ \ \end{array} } & {} \\
\left(\begin{array}{@{} cc:cc @{}} a_{(0,0)}&a_{(1,0)}&b_{(1,1)}&b_{(0,1)}  \\
\tilde{a}_{(1,0)}&\tilde{a}_{(0,0)}& \framebox{$-b_{(0,1)}^{\;\;t}$}&\tilde{b}_{(1,1)}   \\[3mm] \hdashline
c_{(1,1)}&c_{(0,1)}&-a_{(0,0)}^{\;\;t}&-\tilde{a}_{(1,0)}^{\;\;t}   \\
 \framebox{$-c_{(0,1)}^{\;\;t}$}&\tilde{c}_{(1,1)}&-a_{(1,0)}^{\;\;t}&-\tilde{a}_{(0,0)}^{\;\;t} 
\end{array}\right)
 & \hspace{-2mm}
		{\begin{array}{c}      p \\[2mm] n-p \\[2mm] p \\[2mm] n-p  \end{array} }
  \end{array} 
\label{sp}	
\end{equation}
where $b_{(1,1)}$, $\tilde{b}_{(1,1)}$, $c_{(1,1)}$ and $\tilde{c}_{(1,1)}$ are symmetric matrices of appropriate size.
Compare this with the matrices of $\sp(2n)$ given in~\eqref{LA}.
The only difference is a sign difference in the two blocks that are put in a frame in~\eqref{sp}.
We have
\begin{align*}
&\dim \g_{(0,0)}= p^2+(n-p)^2\\
&\dim\g_{(0,1)} = 2p(n-p), \quad \dim\g_{(1,0)} = 2p(n-p)\\ 
&\dim\g_{(1,1)} = p(p+1)+(n-p)(n-p+1) ,
\end{align*}
and obviously by the previous remark: {$\dim \sp_{p}(2n) = \dim \sp(2n) $}.

Having this form, one can verify that {$\sp_{p}(2n)$} consists of all matrices $A$ of $\ssl_{p,n-p,n-p,p}(2n)$ 
that satisfy
\begin{equation}
A^T J +  J A=0
\label{spJ}
\end{equation}
where
\begin{equation}
J= \begin{array}{c c}
\left(\begin{array}{@{} cc:cc @{}} 
  \ 0\  & \ 0\  & \ I\  & \ 0\   \\[1mm]
  0 & 0 & 0 & I \\[1mm]\hdashline
  -I & 0 & 0 & 0   \\[1mm]
  0 & I & 0 & 0 
\end{array}\right)
 & \hspace{-2mm}
		{ \begin{array}{c}      p \\[1mm] n-p \\[1mm] p \\[1mm] n-p  \end{array} }
  \end{array} 
\end{equation}
and $I$ is an identity matrix of appropriate size.
Note that $J^T=-J$ and $J^{-1}=J^t$.
From this, it is easy to show that $\lb A,B\rb$ satisfies \eqref{spJ} when $A$ and $B$ satisfy \eqref{spJ}.

Next, we turn to the class of $\Z^2_2$-GLA's of type $D$. 
The $\Z_2\times\Z_2$-graded LA $\g=\so_{p}(2n)$ consists of all matrices of the following block form:
\begin{equation}
\begin{array}{c c}
    {\begin{array} {@{} c c c c @{}} \ \ \ \ p\ \ & \ n-p\ \ & \ p \ \ \ & \ \ n-p\ \ \end{array} } & {} \\
\left(\begin{array}{@{} cc:cc @{}} 
a_{(0,0)}&a_{(1,0)}&b_{(1,1)}&b_{(0,1)}  \\
\tilde{a}_{(1,0)}&\tilde{a}_{(0,0)}& \framebox{$b_{(0,1)}^{\;\;t}$}&\tilde{b}_{(1,1)}   \\[3mm] \hdashline
c_{(1,1)}&c_{(0,1)}&-a_{(0,0)}^{\;\;t}&-\tilde{a}_{(1,0)}^{\;\;t}   \\
\framebox{$c_{(0,1)}^{\;\;t}$}&\tilde{c}_{(1,1)}&-a_{(1,0)}^{\;\;t}&-\tilde{a}_{(0,0)}^{\;\;t} 
\end{array}\right)
 & \hspace{-2mm}
		{ \begin{array}{c} p \\[2mm] n-p \\[2mm] p \\[2mm] n-p  \end{array} }
  \end{array} 
\label{soD}	
\end{equation}
where $b_{(1,1)}$, $\tilde{b}_{(1,1)}$, $c_{(1,1)}$ and $\tilde{c}_{(1,1)}$ are antisymmetric matrices.
Now one should compare these matrices with the matrices of $\so(2n)$ given in~\eqref{LA}.
Again, the only difference is a sign difference in the two blocks that are put in a frame in~\eqref{soD}.
One can verify
\begin{align*}
&\dim \g_{(0,0)}= p^2+(n-p)^2\\
&\dim\g_{(0,1)} = 2p(n-p), \quad \dim\g_{(1,0)} = 2p(n-p)\\ 
&\dim\g_{(1,1)} = p(p-1)+(n-p)(n-p-1) ,
\end{align*}
and obviously {$\dim \so_{p}(2n) = \dim \so(2n) $}.

There is a similar characterization as before.
One can check that {$\so_{p}(2n)$} consists of all matrices $A$ of $\ssl_{p,n-p,n-p,p}(2n)$ 
that satisfy
\begin{equation}
A^T K +  K A=0
\label{soK}
\end{equation}
where
\begin{equation}
K= \begin{array}{c c}
\left(\begin{array}{@{} cc:cc @{}} 
  \ 0\  & \ 0\  & \ I\  & \ 0\   \\[1mm]
  0 & 0 & 0 & I \\[1mm] \hdashline
  I & 0 & 0 & 0   \\[1mm]
  0 & -I & 0 & 0 
\end{array}\right)
 & \hspace{-2mm}
		{\begin{array}{c} p \\[1mm] n-p \\[1mm] p \\[1mm] n-p  \end{array} }
  \end{array} .
\end{equation}
From the identities $K^T=K$ and $K^{-1}=K^t$ one can again deduce that the graded bracket of $A$ and $B$ satisfies~\eqref{soK} when $A$ and $B$ satisfy~\eqref{soK}.

Finally, the last class of $\Z^2_2$-GLA's listed here is that of type $B$. 
The $\Z_2\times\Z_2$-graded LA {$\g=\so_{p}(2n+1)$} consists of all matrices of the following block form:
\begin{equation}
\begin{array}{c c}
 {\begin{array} {@{} c c c c c @{}} \ \ p\ \ & \ \ \ n-p\ \ & \ \ p \ \ \ & \ \ n-p\ \ & \ 1 \ \ \end{array} } & {} \\
 \left(\begin{array}{@{} cc:cc:c @{}} a_{(0,0)}&a_{(1,1)}&b_{(0,0)}&b_{(1,1)}&c_{(0,1)}  \\[1mm]
       \tilde{a}_{(1,1)}&\tilde{a}_{(0,0)}&b_{(1,1)}^{\;\;t}&\tilde{b}_{(0,0)}&c_{(1,0)}  \\[1mm] \hdashline &&&&\\[-3mm]
       d_{(0,0)}&d_{(1,1)}&-a_{(0,0)}^{\;\;t}&\tilde{a}_{(1,1)}^{\;\;t}&e_{(0,1)}  \\[1mm]
       d_{(1,1)}^{\;\;t}&\tilde{d}_{(0,0)}&a_{(1,1)}^{\;\;t}&-\tilde{a}_{(0,0)}^{\;\;t}&e_{(1,0)}  \\[1mm] \hdashline &&&&\\[-3mm]
      -e_{(0,1)}^{\;\;t}&-e_{(1,0)}^{\;\;t}&-c_{(0,1)}^{\;\;t}&-c_{(1,0)}^{\;\;t}&0  
      \end{array}\right)
 & \hspace{-2mm}
		{\begin{array}{c} p \\[1.5mm] n-p \\[1.5mm] p \\[1.5mm] n-p \\[2mm] 1 \end{array} }
\end{array} 
\label{soB}
\end{equation}
where $b_{(0,0)}$, $\tilde{b}_{(0,0)}$, $d_{(0,0)}$ and $\tilde{d}_{(0,0)}$ are antisymmetric matrices.
There are some other matrix forms for $\g=\so_{p}(2n+1)$, see~\cite{SV2023}.
Note that
\begin{align*}
&\dim \g_{(0,0)}= 2n^2-n-4p(n-p)^2\\
&\dim\g_{(0,1)} = 2p, \quad \dim\g_{(1,0)} = 2(n-p)\\ 
&\dim\g_{(1,1)} = 4p(n-p),
\end{align*}
and $\dim \so_{p}(2n+1) = \dim \so(2n+1)$.
 
Just as for the other classes, there is a different characterization:
{$\g=\so_{p}(2n+1)$} consists of all matrices $A$ of $\ssl_{2p,1,0,2n-2p}(2n+1)$ 
that satisfy
\begin{equation}
 A^T K' +  K' A=0
\label{soK'}
\end{equation}
where
\begin{equation}
K'= \begin{array}{c c}
\left(\begin{array}{@{} cc:cc:c @{}} 
  \ 0\  & \ 0\  & \ I\  & \ 0\ & \ 0\  \\[1mm]
  0 & 0 & 0 & -I & 0 \\[1mm] \hdashline
  I & 0 & 0 & 0  & 0 \\[1mm]
  0 & -I & 0 & 0 & 0 \\[1mm] \hdashline
	0 & 0 & 0 & 0 & 1
\end{array}\right)
 & \hspace{-2mm}
		{\begin{array}{c}      p \\[1mm] n-p \\[1mm] p \\[1mm] n-p \\[1mm] 1  \end{array} }
  \end{array} 
\end{equation}
In this case, $K'^T=K'$ and $K'^{-1}=K'^t$. These relations are again sufficient to show that the matrices of type~\eqref{soB} close under the $\Z_2\times\Z_2$-graded bracket.

\section{Example and conclusions}
\label{sec:4}

A number of examples of systems displaying a $\Z_2\times\Z_2$-graded Lie symmetry were already given in~\cite{SV2023}.
Here, let us again consider an example in the context of parafermion operators.
Ordinary parafermion operators satisfy certain triple commutation relations~\cite{Green,GM}, 
and form a generating set for the Lie algebra $\so(2n+1)$~\cite{KT,Ryan,SV2008}.
Consider now the following set of generators from the $\Z_2^2$-GLA $\so_q(2n+1)$, determined by the last row and last column in~\eqref{soB}:
\begin{align}
&f_{j}^-= \sqrt{2}(e_{j, 2n+1}-e_{2n+1,n+j}), \nonumber\\
&f_{j}^+=\sqrt{2}(e_{2n+1,j}-e_{n+j,2n+1}), \qquad j=1,\ldots , n. 
\end{align}
As usual, $e_{jk}$ is the notation of a matrix of the relevant size 
with zeros everywhere except a 1 on the intersection of row~$j$ and column~$k$.
In terms of these generators, the subspaces of $\so_q(2n+1)$ consist of the following elements:
\begin{align*}
& \g_{(0,1)}=\spn \{ f_{k}^{\pm},\; k=1,\ldots,q \} \\
& \g_{(1,0)}=\spn \{f_{k}^{\pm},\; k=q+1,\ldots,n  \} \\
& \g_{(0,0)}=\spn \{ [ f_{k}^\xi,  f_{l}^\eta],\; \xi, \eta =\pm, \; k,l=1,\ldots,q \; \hbox{and} \; k,l=q+1,\ldots,n   \} \\
& \g_{(1,1)}=\spn \{ \{f_{k}^{\xi}, f_{l}^{\eta}\}, \; \xi, \eta =\pm, \;k= 1,\ldots ,q,\; l=q+1,\ldots n \}.
\end{align*}
The generators $f_i^\pm,\ i=1,\ldots,n$ consists of two sorts of parafermion operators $f_i^\pm,\ i=1,\ldots,q$ and $f_i^\pm,\ i=q+1,\ldots,n$. 
Each sort is subject to the common triple relations~\cite{SV2008} of parafermion statistics
($j,k,l=1,\ldots,q$ or $j,k,l=q+1,\ldots,n$):
\begin{equation}
[[f_{ j}^{\xi}, f_{ k}^{\eta}], f_{l}^{\epsilon}]=\frac12
(\epsilon -\eta)^2
\delta_{kl} f_{j}^{\xi} -\frac12  (\epsilon -\xi)^2
\delta_{jl}f_{k}^{\eta}, \;\; \xi, \eta, \epsilon =\pm\hbox{ or }\pm 1, \label{PF}
\end{equation}
in terms of nested commutators only.
However, the system as a whole is very different from an ordinary set of parafermions.
Indeed, the ``relative commutation relations'' between the two sorts of parafermion operators are as follows, and given only in terms of nested anticommutators 
($j=1,\ldots,q,\ k=q+1,\ldots,n, \ l=1,\ldots,n$ or $j=q+1,\ldots,n,\ k=1,\ldots,q, \ l=1,\ldots,n$):
\begin{equation}
\{\{ f_{ j}^{\xi}, f_{ k}^{\eta}\} , f_{l}^{\epsilon}\} =\frac12
(\epsilon -\eta)^2
\delta_{kl} f_{j}^{\xi} +\frac12  (\epsilon -\xi)^2
\delta_{jl}f_{k}^{\eta}, \;\; \xi, \eta, \epsilon =\pm\hbox{ or }\pm 1. \label{PFrel}
\end{equation}
This gives a new type of parastatistics, not considered before and worth studying.
It could describe a model consisting of two sets of parafermionic particles,
where the two sets are not independent of each other but there is a kind of entanglement determined by~\eqref{PFrel}.

%
%
%
%

\end{document}